 \documentclass[acmsmall,screen]{acmart}
\AtBeginDocument{%
  \providecommand\BibTeX{{%
    \normalfont B\kern-0.5em{\scshape i\kern-0.25em b}\kern-0.8em\TeX}}}
\usepackage{enumitem}

\setcopyright{acmlicensed}
\copyrightyear{2024}
\acmYear{2024}
\acmDOI{XXXXXXX.XXXXXXX}

\acmISBN{978-1-4503-XXXX-X/18/06}

\acmJournal{TOSEM}
\acmVolume{0}
\acmNumber{0}
\acmArticle{0}

\usepackage{color}
\usepackage{balance}
\definecolor{lightgray}{gray}{0.85}

\usepackage{todonotes}
\setuptodonotes{inline}

\renewenvironment{quotation}{%
   \list{}{%
     \leftmargin 0.3cm   
     \rightmargin\leftmargin
   }
   \item\relax
}
{\endlist}

\newcommand\greybox[1]{%
  \vskip\baselineskip%
  \par\noindent\colorbox{lightgray}{%
    \begin{minipage}{.97\columnwidth}#1\end{minipage}%
  }%
  \vskip\baselineskip%
}

\begin{document}

\title{Making Software Development More Diverse and Inclusive: Key Themes, Challenges, and Future Directions}



\author{Sonja M. Hyrynsalmi}
\orcid{0000-0002-1715-6250}
\affiliation{%
  \institution{LUT University}
  \city{Lahti}
  \country{Finland}}
\email{sonja.hyrynsalmi@lut.fi}

\author{Sebastian Baltes}
\orcid{0000-0002-2442-7522}
\affiliation{%
  \institution{University of Bayreuth}
  \city{Bayreuth}
  \country{Germany}}
\email{sebastian.baltes@uni-bayreuth.de}

\author{Chris Brown}
\orcid{0000-0002-6036-4733}
\affiliation{%
  \institution{Virginia Tech}
  \city{Blacksburg, VA}
  \country{USA}}
\email{dcbrown@vt.edu}

\author{Rafael Prikladnicki}
\orcid{0000-0003-3351-4916}
\affiliation{%
  \institution{PUCRS University}
  \city{Porto Alegre}
  \country{Brazil}}
\email{rafaelp@pucrs.br}

\author{Gema Rodriguez-Perez}
\orcid{0000-0002-0062-8418}
\affiliation{%
  \institution{University of British Columbia}
  \city{Kelowna, BC}
  \country{Canada}}
\email{gema.rodriguezperez@ubc.ca}

\author{Alexander Serebrenik}
\orcid{0000-0002-1418-0095}
\affiliation{%
  \institution{Eindhoven University of Technology}
  \city{Eindhoven}
  \country{Netherlands}}
\email{a.serebrenik@tue.nl}

\author{Jocelyn Simmonds}
\orcid{0000-0002-1253-9260}
\affiliation{%
  \institution{University of Chile}
  \city{Santiago}
  \country{Chile}}
\email{jsimmond@dcc.uchile.cl}

\author{Bianca Trinkenreich}
\orcid{0000-0001-7302-6082}
\affiliation{%
 \institution{Colorado State University}
  \city{Fort Collins, Colorado}
  \country{USA}}
\email{Bianca.trinkenreich@colostate.edu}

\author{Yi Wang}
\orcid{0000-0003-1321-4035}
\affiliation{%
  \institution{Beijing University of Posts and Telecommunications}
  \city{Beijing}
  \country{China}}
\email{wang@cocolabs.org}

\author{Grischa Liebel}
\orcid{0000-0002-3884-815X}
\affiliation{%
 \institution{Reykjavik University}
  \city{Reykjavik}
  \country{Iceland}}
\email{grischal@ru.is}

\renewcommand{\shortauthors}{Hyrynsalmi et al.}

\begin{abstract}
\textbf{Introduction}: Digital products increasingly reshape industries, influencing human behavior and decision-making. However, the software development teams developing these systems often lack diversity, which may lead to designs that overlook the needs, equal treatment or safety of diverse user groups. These risks highlight the need for fostering diversity and inclusion in software development to create safer, more equitable technology.  


\textbf{Method}: This research is based on insights from an academic meeting in June 2023 involving 23 software engineering researchers and practitioners. We used the collaborative discussion method 1-2-4-ALL as a systematic research approach and identified six themes around the theme ``challenges and opportunities to improve Software Developer Diversity and Inclusion (SDDI)''. We identified benefits, harms, and future research directions for the four main themes. Then, we discuss the remaining two themes, Artificial Intelligence \& SDDI and AI \& Computer Science education, which have a cross-cutting effect on the other themes.  

\textbf{Results}: This research explores the key challenges and research opportunities for promoting SDDI, providing a roadmap to guide both researchers and practitioners. We underline that research around SDDI requires a constant focus on maximizing benefits while minimizing harms, especially to vulnerable groups. As a research community, we must strike this balance in a responsible way.

\end{abstract}



\maketitle

\section{Introduction}


Software systems are ubiquitous in society, impacting nearly all aspects of modern life as a growing number of companies and industries across domains rely on and deliver products as digital services~\cite{andreessen2011software}. For instance, recent studies show approximately 5.45 billion people, 67.1\% of the global population, rely on the internet---representing a ``supermajority'' of the planet and still rapidly increasing around the world~\cite{datareportal}. The increasing dependence on software has led to the digitization of agriculture~\cite{bHogel2015will}, food sourcing~\cite{paavola2020algorithmic}, environmental monitoring~\cite{hino2018machine}, legal proceedings~\cite{zhurkina2021digitalization}, and many other traditionally non-digitized domains. Software engineering (SE), that is, the processes, methods, and tools to support the development and maintenance of software~\cite{pressman2005software}, is crucial for producing high-quality applications that increasingly impact human behavior, well-being, and decision-making. Recent innovations--such as the advent of generative artificial intelligence (AI) and large language models (LLMs) and machine learning-based systems--have transformed the software development landscape and introduced novel approaches to automate and support SE tasks~\cite{de2024fine,white2023chatgpt}.
For example, GitHub states that, as of February 2023, their coding assistant Copilot is ``behind an average of 46\% of developers' code across all programming languages''\footnote{\url{https://github.blog/2023-02-14-github-copilot-now-has-a-better-ai-model-and-new-capabilities/}}.

Generative AI is experiencing a global surge in software development activities, with the number of relevant open source projects on GitHub more than doubling in 2022---increasing by 248\% with individual contributors to these projects up 148\% around the world~\cite{octoverse}. However, despite recent technological innovations, SE is largely dependent on the efforts of software developers. For instance, prior work concludes ``\textit{significant human involvement and expertise}''~\cite[p.~13]{white2023chatgpt} are necessary for leveraging LLMs to automate tasks related to the design, implementation, testing, and maintenance of applications, and there is a call for generative {AI} in software engineering to be human-centered~\cite{Manifesto2024}. 
However, the diversity and values of software development teams that design software systems often do not reflect the diversity and values of intended users, or, more broadly, our society. For example, the European Institute for Gender Equality estimates that only 17\% of ICT specialists in Europe are women\footnote{\url{https://eige.europa.eu/publications-resources/toolkits-guides/work-life-balance/women-in-ict}}. Similarly, in the 2022 Stack Overflow Developer Survey, approximately 92\% of the respondents identified as male and 77\% identified as White or European~\cite{so_survey}\footnote{More recent versions of this survey, deployed in 2023 and 2024, exclude these demographic data.}. In contrast, the global population is approximately 50\% male~\cite{statista} and 16\% White~\cite{worldnotwhite}. 

This ``diversity crisis''~\cite{albusays2021diversity} in SE can have major ramifications for people from underrepresented backgrounds. For example, research shows that most software lacks gender inclusivity, favoring the problem-solving processes of men~\cite{gendermag}. Moreover, this crisis contributes to non-inclusive environments where minority developers are disadvantaged. For example, peer code reviews, where developers review code from contributors before merging into source code, are a common practice to improve software quality~\cite{mcintosh2016empirical}. However, recent studies show that code contributions from programmers of non-White and non-male backgrounds receive more pushback~\cite{murphy2022pushback} and higher rejection rates~\cite{nadri2021insights,terrell2017gender}.

Research reveals that diversity and inclusion enhance SE~\cite{conversation,tech2019workforce}. For example, studies show that open source projects with diverse contributors are more productive~\cite{vasilescu2015gender}, neuroinclusive teams are more productive than purely neurotypical teams~\cite{CompetitiveAdvantage}, heterogeneous collaboration based on race and ethnicity leads to a higher number of contributions to open source projects~\cite{shameer2023relationship}, and working on gender-diverse teams improves attitudes towards women and improves decision-making and innovation~\cite{kohl2022benefits}. However, a significant amount of work must be done to create inclusive work environments that lead to a more diverse community, which is building the software that is the foundation of our digital society.

However, recently there have been reports of many companies scaling back in their DEI (Diversity, Equity and Inclusion) activities or teams~\cite{burnett2024prevent, alfonseca2023corporate}. Some organizations promoting DEI in the tech industry, such as Girls in Tech and Women Who Code, have also even been forced to shut down due to reduced financial support~\cite{nix2024movement}. While backlash and critical discussions around DEI initiatives have gained significant attention, especially in the media, there are also valid concerns about improving their effectiveness, accountability, and long-term impact~\cite{follmer2024under, robb2024cutting, cox2024counteracting}. At the same time, evidence supporting the business case and the importance of DEI initiatives is growing stronger~\cite{mckinsey2023diversity}. 

Hence, there is a clear mismatch between the existing evidence that finds various benefits of diversity and inclusion in SE, and the work highlighting the rampant diversity crisis in the same field. This mismatch suggests that research findings that underline the benefits of diversity and inclusion might not be strong enough or might not be successfully transferred to industrial practice and that a more detailed understanding of the various factors contributing to the diversity crisis in SE is necessary.

To this end, this paper shares insights from SE researchers and practitioners on the challenges and opportunities regarding diversity and inclusion and creates a roadmap for avoiding undesirable outcomes (``dystopia'') and working towards a desirable state (``utopia''). Our overarching research question is:

\begin{quote}
\textbf{RQ:} \emph{What are the key challenges and research opportunities for promoting Software Developer Diversity and Inclusion (SDDI)?}
\end{quote}

We begin by sketching four contrasting scenarios for SE in 2030: two outlining a desirable state (``utopian future'') and two outlining an undesirable state (``dystopian future''). For these scenarios, we used six themes systematically developed from an academic meeting held in June 2023.
We then proceeded to introduce the four main themes that guided the results section of this paper and stood out as clear, separate themes: methodologies and metrics, intersectionality, knowledge transfer, and socioeconomic understanding in SE. Based on these themes, we offer guidance for researchers to bridge gaps in SDDI.

Finally, two of our original themes, AI \& SDDI and AI \& CS Education, were cross-cutting themes for the four main themes, as they have potential implications for all other themes. For these, we discuss how the rapid developments in AI can challenge or support progress in SDDI. We also underline how a carefully balanced use of AI is essential for building an inclusive future that avoids undesirable scenarios and gets as close as possible to a desirable future.

\section{Research Process}

The results presented in this paper originate from facilitated discussions during an academic meeting in June 2023\footnote{\url{https://shonan.nii.ac.jp/seminars/194}}. This meeting brought together 23 software engineering researchers and practitioners from diverse backgrounds and career phases interested in fostering SDDI-related research. 18 workshop participants were actively conducting SE research at academic institutions from around the world, while five were researchers and/or developers at companies in industry. The workshop was organized in a hybrid format, with seven participants joining remotely. The organizers of the meeting followed NII Shonan's instructions for balanced participation, which are stated as follows: \textit{``Please make a good balance of international diversity, research fields, and a proportion of theoretical and practical areas. The list also should be a mixture of proven experts and promising young researchers, preferably including representative researchers of their community.''}. All the participants, from PhD candidates to full professors, were experienced in working with SDDI topics. 

    \begin{figure}
    \centering
    \includegraphics[width=0.8\textwidth]{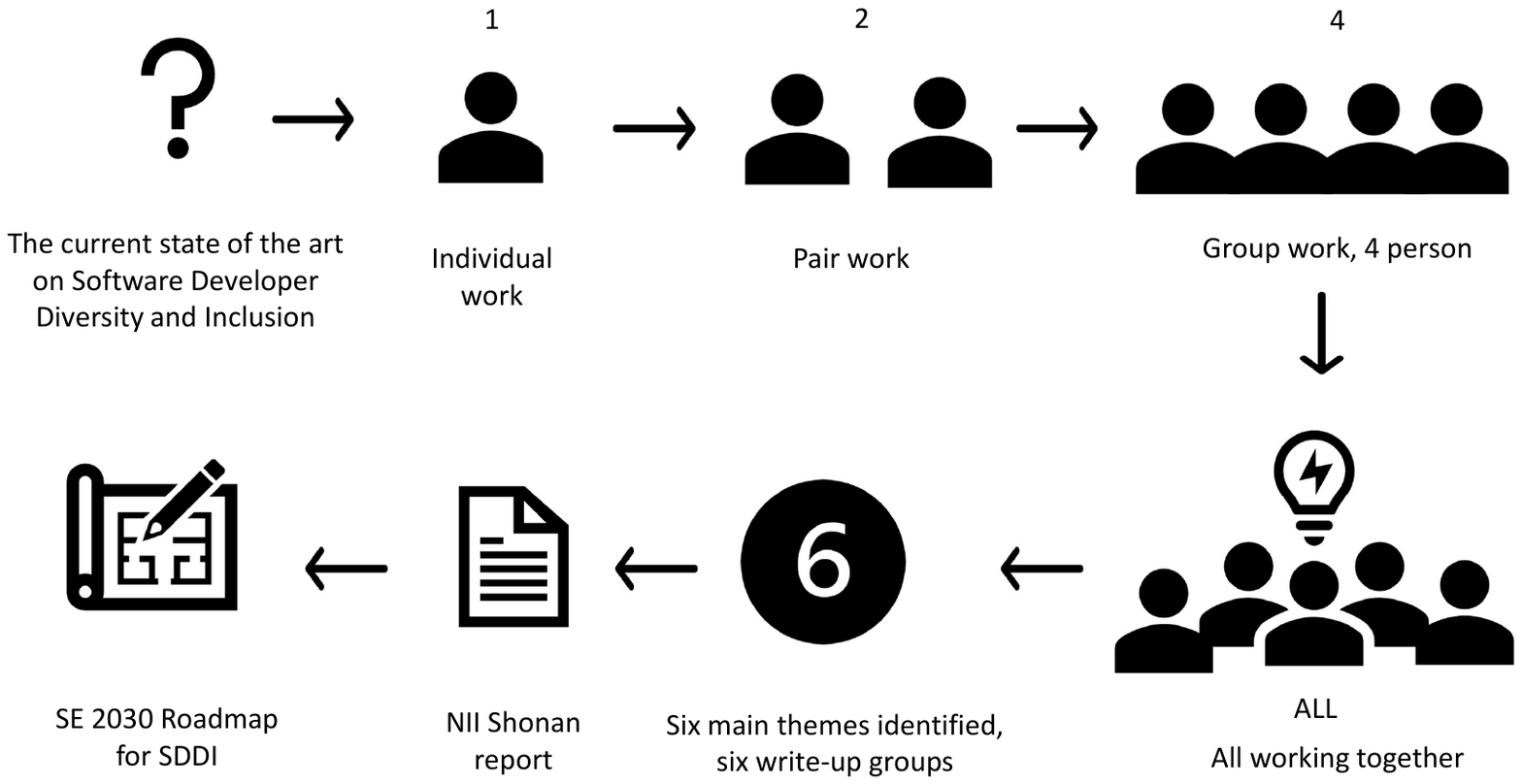}
    \caption{Overview of our 1-2-4-ALL process for understanding opportunities for SDDI through SE Research}
    \label{fig: 124all}
    \end{figure}

We utilized 1-2-4-ALL as a systematic data collection method to allow self-reflection and collaborative discussion, building toward consensus or shared understanding among participants and to foster inclusive participation and generate a wide range of ideas from the participants in the meeting~\cite{lipmanowicz2013surprising}\footnote{\url{https://www.liberatingstructures.com/1-1-2-4-all/}}. An overview of our method is presented in Figure \ref{fig: 124all}. The 1-2-4-ALL method is one of the Liberating structures methods, which are techniques to support lively discussions and foster engagement in a group setting. Other popular liberating structures methods are, e.g., Impromptu Networking and Crowd Sourcing~\cite{liberating}. Liberating structures can be referred to as a ``destructive methodology'', as it tries to create a discussion and communication between participants without classic ways for one-way communication such as PowerPoint presentations~\cite{ferguson2015applying}. 

For this study, we decided to use 1-2-4-ALL as it is a more structured way to approach shared decisions than, for example, the Impromptu Networking method, which is more spontaneous
structure for sharing of ideas and engage the participants~\cite{ferguson2015applying}. Prior work has leveraged the 1-2-4-ALL method to explore techniques to improve student engagement in health and science education~\cite{ferguson2015applying}, identify research priorities in psychiatric-mental health~\cite{mahoney2016using}, and enhance behavior and performance for management teams~\cite{kjellstrom2022fostering}. In 1-2-4-ALL, ideation starts with an individual reflection (1 minute), followed by a discussion in pairs, building on ideas from self-reflection (2 minutes),  sharing ideas in groups of four focusing on similarities and differences (4 minutes) and then sharing the best group idea with all the participants. In SE, liberating structures have been used especially as a discussion and decision tool in Agile teams or to ensure that everyone gets their voice heard in remote meetings~\cite{turner2021ebook, kang2023innovations}. 

As the results of the final group discussion, the groups were asked to present one goal, two outcomes, and four themes related to \textit{challenges and opportunities to improve SDDI}. The results of the breakout group discussions were shared with all participants to obtain feedback and reach an agreement on a defined list of SDDI themes. The participants were then divided into groups according to the defined themes to further expand on the research challenges and solutions. After intensive discussions around the identified goals, outcomes, and themes in multiple group sessions, we developed a report\footnote{\url{https://shonan.nii.ac.jp/docs/No\%20.194.pdf}} summarizing the discussion for each theme. 

\begin{figure}
    \centering
    \includegraphics[width=\textwidth]{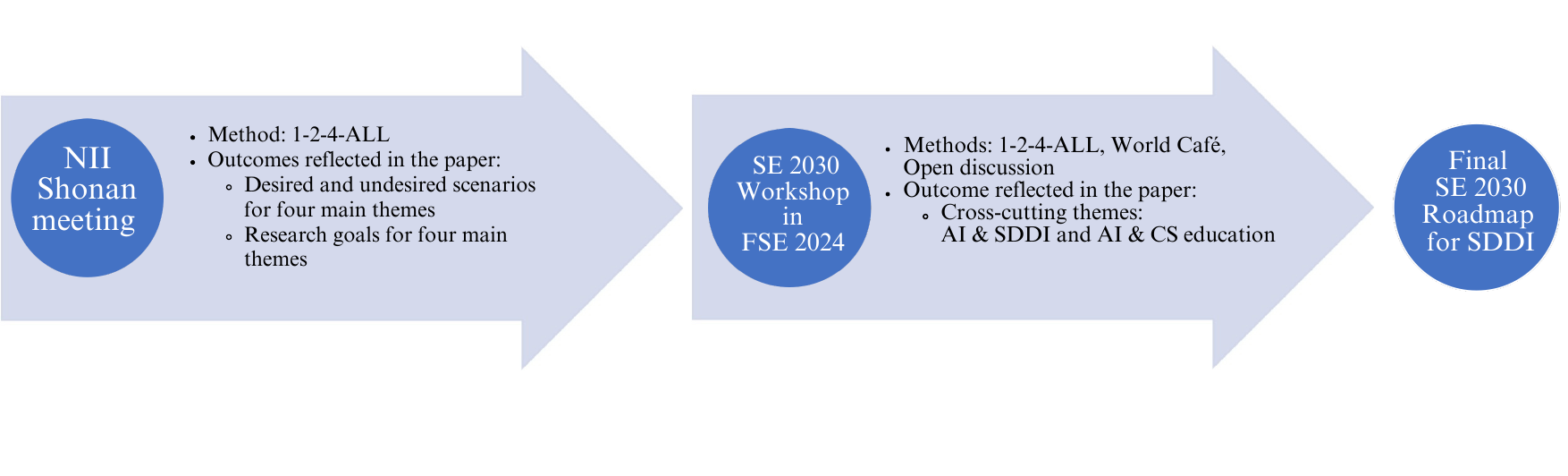}
    \caption{Overview of the process for creating the SE 2030 Roadmap for SDDI. The process started from NII Shonan meeting, continued for the first version presented in SE 2030 Workshop and insights from there led to the final roadmap.}
    \label{fig:process}
    \end{figure}

In this paper, we expand our results from the report to provide a more coherent vision of future research directions for SDDI. We expand the discussion around four of the main themes: methodologies and metrics, intersectionality, knowledge transfer, and connections to socioeconomic understanding. These themes were chosen based on the timeliness, potential impact, and interest participants attributed to the themes. From each of the themes, we identified potential benefits and harms. In turn, the themes, related research literature and the discussion in the SE 2030 workshop held at FSE Conference 2024, where we presented our preliminary results, guided us to form research goals for 2030 (Figure \ref{fig:process}). In the discussion section, we discuss the two remaining themes presented in the original meeting: `AI \& SDDI' and `AI \& CS Education', as these themes were noticed to be cross-cutting.

\section{Four Scenarios for 2030}
\label{sec:scenarios}
In the following, we present two desirable ``utopian'' scenarios, that is, scenarios ``having the characteristics and organization of a perfect society''\footnote{\url{https://dictionary.cambridge.org/dictionary/english/utopian}} and two contrasting undesirable ``dystopian'' scenarios, i.e., scenarios ``relating to a very bad or unfair society in which there is a lot of suffering''\footnote{\url{https://dictionary.cambridge.org/dictionary/english/dystopian}}.
These scenarios capture the future of software development work and the future of education, two areas that we, as SE researchers, can actively study and shape. 
This approach of describing the potential impact of AI in the future in terms of desirable (``utopian'') and undesirable (``dystopian'') outcomes has been used in related work, e.g., \cite{Aquino2023,Chan2023}

\begin{quotation}
\textbf{S1: Desirable Scenario in Workplace:} \emph{Jamie works as a recently-graduated junior software engineer at an up-and-coming tech company. The company is at the forefront of inclusion, offering, e.g., flexible work arrangements that consider factors such as child care responsibilities, support for various career paths and the age of employees and applicants, and preferences concerning work and communication modes. These initiatives are paying off, as the company has successfully attracted a diverse workforce, leading to higher productivity, a welcoming and safe work environment, and more innovative and successful products.
Despite coming from a minority group and being relatively junior, Jamie feels that the company provides a safe space for developing their career and competencies and invites them to participate in decision-making. For instance, the company offers automated tools to support various workflows, e.g., in code review and in programming, that allow Jamie to receive early feedback on their work, spot mistakes, and improve their skills without fear of discrimination. Similarly, the tools and practices at the company allow diverse teams to communicate effectively and positively. The company fosters a culture that encourages the active use and development of these tools for everyone's benefit, both within and beyond the company. In addition, the company carefully monitors existing diversity metrics and plans interventions based on them. Regular surveys help company leadership to consider employees' values in their strategic decisions. Finally, the company employs mechanisms at various stages to avoid potential misuse of data obtained from its employees, e.g., through transparency in decision-making.}
\end{quotation}

Before graduation, Jamie studied at a university where they attended courses taught by Kris:

\begin{quotation}
\textbf{S2: Desirable Scenario in Education:} \emph{Kris is a university-level educator in software engineering. Due to inclusive and equitable conditions at the university, students from varied backgrounds are enrolled in Kris' courses, contributing to a truly diverse and inclusive study experience. Although classes are large so that the university can cope with the continuing strong demand for software engineering professionals, technology advances and government incentives make it possible for all students to receive targeted, individual feedback, regardless of diverse aspects such as gender, nationality, or age. In part, this is achieved through tools and indicators that enable automated feedback and coaching, so that Kris and their staff can fully focus on providing the best possible learning experience. Finally, the assessment has changed from standardized and unified exams and assignments to a personalized form, tailored to suit each student's individual needs. In this environment, students feel safe to make mistakes and express themselves. In turn, this allows fruitful collaborations between students and staff.}
\end{quotation}

To contrast the experiences of Jamie and Kris, we outline an undesirable future in which Ash struggles in their workplace, and Moss struggles in their role as a student.

\begin{quotation}
\textbf{S3: Undesirable Scenario in Workplace:} 
\emph{Ash has graduated from a prestigious university in their home country with a degree in software engineering. They are employed at a local IT company. The company provides software development services to large corporations in rich and highly developed countries. 
Due to the economic realities in their home country, Ash and their company cannot afford to pay for existing top-of-the-shelf tools, such as coding coaches or LLM-based code generation tools. Even for free tools, unreliable and slow internet connections limit the use of these tools. This results in more manual work and less time to develop skills.
Similarly, language barriers prevent them from learning and improving their skills at the same pace as similarly qualified graduates in richer and more developed countries, as existing tools cater only to English speakers. At Ash's employer, traditional gender role models and hierarchical structures persist and affect career progression chances.
As a result, Ash feels that their chance of a successful career is relatively low due to factors outside of their control. An overall lack of awareness of diversity matters also impedes the services provided by Ash's employer, as stereotypes and biases are ingrained in the products provided by the company.}
\end{quotation}

Meanwhile, not only are workers such as Ash in emerging countries struggling, but education in rich and highly developed countries is also not what it used to be: 

\begin{quotation}
\textbf{S4: Undesirable Scenario in Education:} \emph{Moss is a student in software engineering at a local university. Many tasks and activities that used to involve human interaction have been replaced by automated tools. This has made education cheaper for the university. However, instead of lowering tuition fees, student numbers have increased dramatically, but staff have been reduced. 
Lecture topics and assignments lack personalization and typically cover generic and stereotypical examples.
As a result, Moss does not feel like they belong to the university, having almost no interaction with students and staff who come from a similar background as them. As someone who moved to the city for their studies, social integration has therefore been lacking. Instead, Moss feels that education could just as well be remote. 
}
\end{quotation}

\section{Results: Avoiding Undesirable Scenarios}\label{sec:results}
In Section~\ref{sec:scenarios}, we have outlined four scenarios on how SE industrial practice and higher education could look in 2030.
These scenarios do not have a direct connection to SE research.
However, SE research can and should contribute to industrial practice and higher education in order to avoid the undesirable scenarios and instead achieve a future closer to the desirable scenarios we described.
To do so, we outline four orthogonal areas in which we believe SE research needs to evolve in the coming years.

In Scenario S1, several SDDI metrics guided the company's interventions toward a safer place for Jamie to develop their career. 
Similarly, Kris used various indicators in their educational setting in S2.
If the university in Scenario S4 had monitored their students' experience and well-being using suitable metrics, they would have noticed the lack of belonging and feeling of inclusion among students such as Moss.
These three scenarios motivate the use of diversity metrics that can be used by practitioners and educators to better understand diversity in their environment and make decisions based on them. 
These metrics could range from obtaining demographic data, such as gender, country or region of birth, or age, to using various instruments to consider aspects such as personal preferences and interests, or cognitive diversity. Gender, in particular, has been one of the most studied aspects of diversity in software engineering, but the range of diversity aspects in the field has expanded recently~\cite{rodriguez2021perceived}. 
SE research can contribute to this area by developing appropriate \emph{Methodologies and Metrics}.

In Scenario S3, Ash's experiences conflated frustrations due to multiple diversity dimensions: socioeconomic background, language background, and gender.
In contrast, various intersectional backgrounds were effectively supported in Scenario S2 through personalized learning, which has been shown to reduce educational inequality and foster inclusion for learners from diverse cultures~\cite{strekalova2021meeting}.
These scenarios highlight the importance of developing a better understanding of \emph{Intersectionality} in SE research \cite{cech2022intersectional}, that is, when multiple diversity aspects overlap.

Both Scenarios S1 and S3 demonstrate the importance of transferring knowledge about SDDI to and from the SE field. 
For example, in Scenario S3, a lack of transfer of new ideas from outside means that traditional gender roles dominate the company culture, impacting Ash's sense of belonging as they do not identify with these traditional gender roles and as their career progression is affected. Research has shown that women are still underrepresented in the industry, they still leave the industry as they do not feel a sense of belonging or are mistreated, or they even enter the field later in life as they lack guidance and support in choosing IT careers at a younger age.~\cite{oliveira2024navigating, mckinsey2023women, hyrynsalmi2024second}.
In Scenario S1, Jamie and their employer thrive due to the adoption of diversity initiatives. 
This aligns with the findings that a diversity perspective is often missing in innovation and entrepreneurship policies, which predominantly target male-dominated industries. As such, there needs to be an increased push towards \emph{Knowledge Transfer} between SE research, industry, education, government bodies, and societal groups to ensure diversity initiatives are effectively integrated into innovation policies and practices.~\cite{liberda2017innovation}

Finally, industrial practice and higher education do not operate in isolation but are embedded in the larger socioeconomic context.
For example, in Scenarios S1 and S3, factors like income or government funding had a decisive effect on the workplace, highlighting differences in how investments shape outcomes in high- and low-income regions, as research has shown~\cite{okoye2022impact}. 
Therefore, it is imperative to use better existing \emph{Large-scale Socioeconomic Understandings} developed by other disciplines in SE research to provide more inclusive workplace environments and higher education environments.

In the following subsections, we discuss these four themes in more depth and provide future-oriented research goals for each of them. These themes represent important research opportunities for the next five years to bridge the gaps in SDDI in practice.


\subsection{Methodologies and Metrics}
Methodologies and metrics are essential for scientific exploration and discovery \cite{creswell2017research}.
Methodologies entail systematic research approaches for acquiring new knowledge through various research methods for data collection and analysis \cite{wohlin2021guiding}. Metrics are used to quantify a characteristic of a real-world entity \cite{basili1988models}. SE research, in particular, uses a variety of methods and metrics to provide empirical evidence to support software development processes and tools~\cite{easterbrook2008selecting}. For example, previous work has used different techniques and measures to explore concepts related to SDDI, including surveys to understand developers' sense of belonging~\cite{trinkenreich2023belong,trinkenreich2024unraveling}, interviews to understand barriers for women in online programming communities~\cite{ford2016paradise}, and studies mining GitHub repositories to explore toxicity in open source software~\cite{miller2022toxic}. However, while all these examples are related to SDDI, an underlying theory describing how they connect for SDDI and what is the consequence for the software being produced is missing. For example, how do we investigate toxicity as a barrier to minorities' participation and belonging to a software development project? ``How'' includes methods and metrics.


The Linux Foundation’s Community Health Analytics in Open Source Software (CHAOSS)\footnote{\url{https://chaoss.community/unveiling-the-impact-dei-metrics-overcoming-social-barriers-in-open-source/}} has identified several key metrics to enhance diversity. For example: (1) metrics for diversifying recruitment which focus on community engagement and labor investment, promoting equal opportunities and assessing candidates' value; (2) metrics for leadership, which serve as key performance indicators, enhancing diversity and guiding strategies for inclusive participation while preventing authoritarian dynamics; (3) metrics on psychological safety and burnout which create environments where participants can raise their concerns and receive support; (4) mentorship and retention metrics which encourage collaboration and track new member engagement.

Table~\ref{tab:methodologies} summarizes potential benefits, harms and research directions for this theme.

The focus of this theme is not only on what has been published in the field of SE around diversity and inclusion and with which methods---but also a look at multiple domains, that is, SE, educational psychology, and management, to investigate more how insights from these fields could be incorporated into the methodologies and metrics used to research SDDI in SE. 



\subsubsection{Methods}
SDDI research often employs mixed methods~\cite{creswell2017research}, utilizing qualitative and quantitative data.
Using this data more efficiently is an important future goal. \textsc{Weighting}, \textsc{timing}, and \textsc{mixing} are aspects that must be considered before planning mixed-methods SDDI research.

\textsc{Weighting} refers to the priority given to qualitative and quantitative data in research \cite{creswell2017research}. The priority can be the same or favor one over another, depending on the researcher's goals and the audience. For instance, if the aim is to advocate for policy changes, quantitative metrics may take precedence to provide compelling evidence to decision-makers. Conversely, if the focus is on fostering a supportive community, qualitative insights might be prioritized to understand personal experiences better.

\textsc{Timing} is about deciding whether to collect data sequentially in phases or concurrently \cite{creswell2017research}. In sequential data collection, either qualitative or quantitative data collection can occur first, depending on the research goal. 
Concurrent data collection involves simultaneously gathering data for qualitative and quantitative analyses, which can be interesting in time-sensitive projects where contacting participants multiple times for data collection is not feasible (e.g., single surveys with open and closed-ended questions) \cite{trinkenreich2022empirical}. 
Previous SDDI mixed-methods research used both concurrent and sequential mixed-methods research. \citet{trinkenreich2022empirical} followed a concurrent mixed-methods research collecting data through a single survey to qualitatively uncover the challenges faced by women in software development teams and segment those challenges across demographics of age, caregiving responsibilities, marital status, and tenure. Examples of SDDI sequential mixed-methods research included surveys and mining software repository studies in varying orders. \citet{vasilescu2015gender} started with a survey on perceptions of team diversity and then mined a software repository to measure how team productivity and turnover are impacted by gender and tenure diversity. Following the opposite order, \citet{prana2021including} started mining software repositories to investigate differences in gender inclusion in projects across geographical regions, followed by a survey aimed at developers from the various regions about factors that can potentially contribute to differences in developer participation based on gender and geography worldwide.

\textsc{Mixing} involves choosing how to integrate or connect qualitative and quantitative data, which can be done during data collection, analysis, or interpretation. 
For example, in a two-phase project, mining software repositories can be followed by a gender inference approach to support the selection of women's data for a subsequent survey, connecting the two phases, as done by Qiu et al.~\cite{qiu2019going}. In some cases, one form of data may support another, \textit{embedding} a secondary form within a larger study. For example, researchers might primarily collect quantitative data through surveys measuring employees' stress levels and job satisfaction. To enhance these findings, they could embed qualitative interviews where employees share personal stories and experiences related to their well-being. The researcher may weigh the collection of one type of data while using the other type to provide supplementary information.

\subsubsection{Metrics}
Measuring diversity- and inclusion-related phenomena is a challenging task due to the need to capture complex social constructs, e.g., gender identity, sense of belonging or suitability of technology, being mindful that these constructs are highly likely to be perceived and conceptualized very differently by marginalized populations and recognizing the need to focus on the experiences of these populations~\cite{Ford2024}.

At the level of demographics, gender, and in particular opposition between women and men, is the most studied diversity aspect in software engineering~\cite{rodriguez2021perceived}. However, despite the extensive usage of this construct, acquiring information about the gender identity of study participants—whether in controlled experiments, interviews, and surveys—or individuals contributing to datasets used in data-driven archival studies, such as repository mining, presents significant challenges. Inadequate methods for gathering this information risk both alienating participants and undermining the validity of the scientific findings~\cite{Serebrenik2024}. 
Even more so, given the well-known gender-related differences in information processing~\cite{burnett2016gendermag}, one should ask themselves whether gender is an appropriate construct to study as opposed to one of the information processing facets, that can be reliably assessed by means of a validated instrument~\cite{Hamid2024}. 

Finally, regarding metrics for inclusion, SE research has been advancing on measuring the sense of belonging \cite{trinkenreich2023belong,trinkenreich2024unraveling}, which is the extent to which individuals feel like they belong or fit in a given environment \cite{hagerty1995developing}.  Belongingness is a theoretical concept that is hard to observe directly, but it can be asked through different manifest variables (questions) and grouped on a latent construct. There are different instruments in the literature to measure a sense of belonging. The instrument used to measure belongingness in Open Source Software, for example, was based on the concept of a sense of virtual community \cite{blanchard2007developing} (a community that mainly interacts online) and included questions about feelings of membership as a member of the team, being known by others and knowing who to ask for help, feeling valued and perceiving the team is like home \cite{trinkenreich2023belong}. 

In addition to the methods considered, the actual implementation of these methods for research purposes can be improved. 
On the diversity lens, most of the literature related to minorities in SE and underrepresentation is still focused on gender \cite{canedo2020work,trinkenreich2022women,vanBreukelen2023}, race \cite{rodriguez2021perceived,nadri2021insights}, neurodiversity \cite{morris2015understanding,marquez2024inclusion}. English confidence is a metric to evaluate inclusion for people who are non-native in English \cite{serebrenik2020diversity} and can include multiple questions to include both written and spoken communication, and both technical and social contexts \cite{trinkenreich2023belong}. Socio-economic factors are also essential to be measured. For example, Goel et al. suggest that most research for end-user programming targets WEIRD (Western, Educated, Industrialized, Rich, and Democratic) users, while ignoring non-WEIRD populations that make up 85\% of the world~\cite{goel2023end}.

We need more diversity aspects and intersection of those, which is going to be discussed in the next section. 
%
%
The overarching research goal for this theme is as follows:
\greybox{
\textbf{Research Goal:} Develop methodologies and metrics to effectively analyze diversity and inclusion in software engineering, making use of mixed methods and community data.
}

\begin{table}
\centering
\caption{Methodologies and Metrics: Benefits, Harms, and Research Directions}
\label{tab:methodologies}
\footnotesize
\begin{tabular}{|p{0.3\linewidth}|p{0.3\linewidth}|p{0.3\linewidth}|}
\hline
\textbf{Benefits} & \textbf{Harms} & \textbf{Research Directions} \\
\hline
\begin{itemize}[leftmargin=5pt]
    \item Enhanced inclusion metrics
    \item Comprehensive understanding of community dynamics
    \item Tailored intervention strategies for diversity
\end{itemize} & 
\begin{itemize}[leftmargin=5pt]
    \item The definition of inclusion not always clear
    \item Lack of comprehensive analysis tools
    \item Oversimplified interpretations of complex identities
\end{itemize} & 
\begin{itemize}[leftmargin=5pt]
    \item Bridging qualitative and quantitative research
    \item Developing new theoretical frameworks and theories from the data
    \item Utilizing online community data for inclusiveness measures
\end{itemize} \\
\hline
\end{tabular}
\end{table}

\subsection{Intersectionality}

Introduced by Kimberl{\'e}  Crenshaw~\cite{Crenshaw1988}, the concept of intersectionality describes the ways in which social categories of identity, difference, and disadvantage, e.g., gender, race, ethnicity, sexual orientation,
gender identity, disability, class, age, and other forms of discrimination, ``intersect'' simultaneously to create unique dynamics and effects~\cite{Cole2009}. Intersectionality suggests that different diversity aspects are not mutually exclusive and do not operate in isolation. Research also shows the negative consequences for individuals at the intersection of diversity categories in SE and computing. For example, Ross et al. show that fewer black women are
introduced to CS than non-Black women and Black men~\cite{Ross2020}. In this case, Black women do not know whether their negative
experiences should be attributed to their gender or race. Similarly, a recent study by van Breukelen et al.~\cite{vanBreukelen2023} shows that older women developers adopt various ``survival strategies'' to persist in the tech industry, and are uncertain whether their negative
experiences in software development environments are due to ageism~\cite{baltes2020age} or sexism. 

Black women and older women are merely two possible intersections to consider when studying SDDI from an intersectional perspective. Many individuals also find themselves at the intersection of more than two diversity axes. Research suggests that White, able-bodied, and heterosexual male STEM professionals experience favourable treatment, while people with more intersections face reduced social inclusion, professional respect, career opportunities, salaries, and persistent intentions~\cite{cech2022intersectional}. Therefore, more work is needed to understand the experiences of developers at the intersection of multiple diversity aspects. 
While this need has been recognized, the number of such studies remains limited~\cite{ford2019remote,Sanchez-Gordon2024,Kohl2024,Szlavi2023}.

Table~\ref{tab:intersectionality} presents benefits, harms and research directions on intersectionality in SE. The benefits and harms clearly show the complexity of the theme and the dangers of ignoring intersectionality.
Research directions involve improving measurements of bias, investigating the experiences of individuals with diverse identities not or under-explored in SE literature so far, considering the experiences of individuals across diversity axes, and designing interventions and guidelines to support developers who identify with multiple
diversity aspects. The overarching research goal for this theme is as follows:
\greybox{
\textbf{Research Goal:} Understand challenges and motivate solutions to support developers who identify with multiple marginalized groups.
}

\begin{table}[htbp]
\centering
\caption{Intersectionality and SE: Benefits, Harms, and Research Directions}
\label{intersectionality}
\footnotesize
\begin{tabular}{|p{0.3\linewidth}|p{0.3\linewidth}|p{0.3\linewidth}|}
\hline
\textbf{Benefits} & \textbf{Harms} & \textbf{Research Directions} \\
\hline
\begin{itemize}[leftmargin=5pt]
    \item In-depth understanding of diverse identities
    \item Enhanced methodologies for capturing intersectionality
    \item Empowerment through tailored interventions
\end{itemize} & 
\begin{itemize}[leftmargin=5pt]
    \item Potential for oversimplification in the analysis
    \item Danger of marginalization through one-size-fits-all policies and analysis
    \item Ethical concerns in data collection and analysis
\end{itemize} & 
\begin{itemize}[leftmargin=5pt]
    \item Developing a two-stage research approach to utilize both qualitative and quantitative data
    \item Understanding underexplored intersections of diversity in SE
    \item Designing interventions that respect and enhance self-perception and self-efficacy
    \item Adapting to AI-powered development and research environments
\end{itemize} \\
\hline
\end{tabular}
\label{tab:intersectionality}
\end{table}

\subsection{Knowledge Transfer}

Substantial research activity is already ongoing in SE and beyond, targeting the effect of various diversity dimensions on the workforce. For example, existing work shows that masculine cultures can alienate women developers~\cite{faulkner2007nuts,GirlsWhoCode}---yet, studies show increased inclusion of women in development teams can enhance productivity~\cite{vasilescu2015gender}, community~\cite{catolino2019gender} and code quality~\cite{terrell2017gender}. However, transferring these findings to other actors is ultimately vital to enhancing software development and quality. Overall, co-creation or transferring findings and best practices between different actors such as industry, academia, government, and society is becoming more crucial, as technology has more cross-cutting impact in every sector. Therefore, we approach the topic of knowledge transfer in advancing SDDI.
Table~\ref{knowledge_transfer} summarizes the benefits of successful knowledge transfer, the harms of ignoring knowledge transfer, and research directions for the coming years.

\begin{table}
\centering
\caption{Knowledge Transfer and SDDI: Benefits, Harms, and Research Directions}
\label{knowledge_transfer}
\footnotesize
\begin{tabular}{|p{0.3\linewidth}|p{0.3\linewidth}|p{0.3\linewidth}|}
\hline
\textbf{Benefits} & \textbf{Harms} & \textbf{Research Directions} \\
\hline
\begin{itemize}[leftmargin=5pt]
    \item Seamless integration of research into industry practices
    \item Enhanced innovation through collaborative efforts
    \item Better alignment of academic curriculum with industry needs
\end{itemize} & 
\begin{itemize}[leftmargin=5pt]
    \item Fragmented and siloed knowledge pools
    \item Industry and educational practices disconnected from current research and vice versa in the area of SDDI
    \item Overlap in the actions, actions not visible to others
\end{itemize} & 
\begin{itemize}[leftmargin=5pt]
    \item Establishing frameworks for continuous exchange between academia and industry
    \item Cultivating partnerships for mutual knowledge enhancement
    \item Recognizing key actors from all areas of the quadruple helix
\end{itemize} \\
\hline
\end{tabular}
\end{table}

In the last decades, SE research has increasingly paid attention to showing industrial relevance in published work.
This is witnessed by an increasing amount of publications with joint academic and industrial authors, special forums for industry-relevant work, such as the SE in Practice track at the International Conference on Software Engineering (ICSE),\footnote{\url{https://conf.researchr.org/track/icse-2025/icse-2025-software-engineering-in-practice}} a special issue in IEEE Software on sustaining software engineering knowledge transfer,\footnote{\url{https://www.computer.org/digital-library/magazines/so/call-for-papers-special-issue-on-sustaining-software-engineering-knowledge-transfer}} or funding calls that require collaboration between academia and industry \cite{BusinessFinland, LinkageProjects}.
This focus on industry-relevant research has led to substantial work on transferring technology and knowledge, typically from academia to industry (see, e.g., \cite{gorschek2006model}).
However, we explicitly question whether the transfer of SE research knowledge, especially, has indeed been successful. 

\begin{figure}[htbp]
    \centering
    \includegraphics[width=0.5\textwidth]{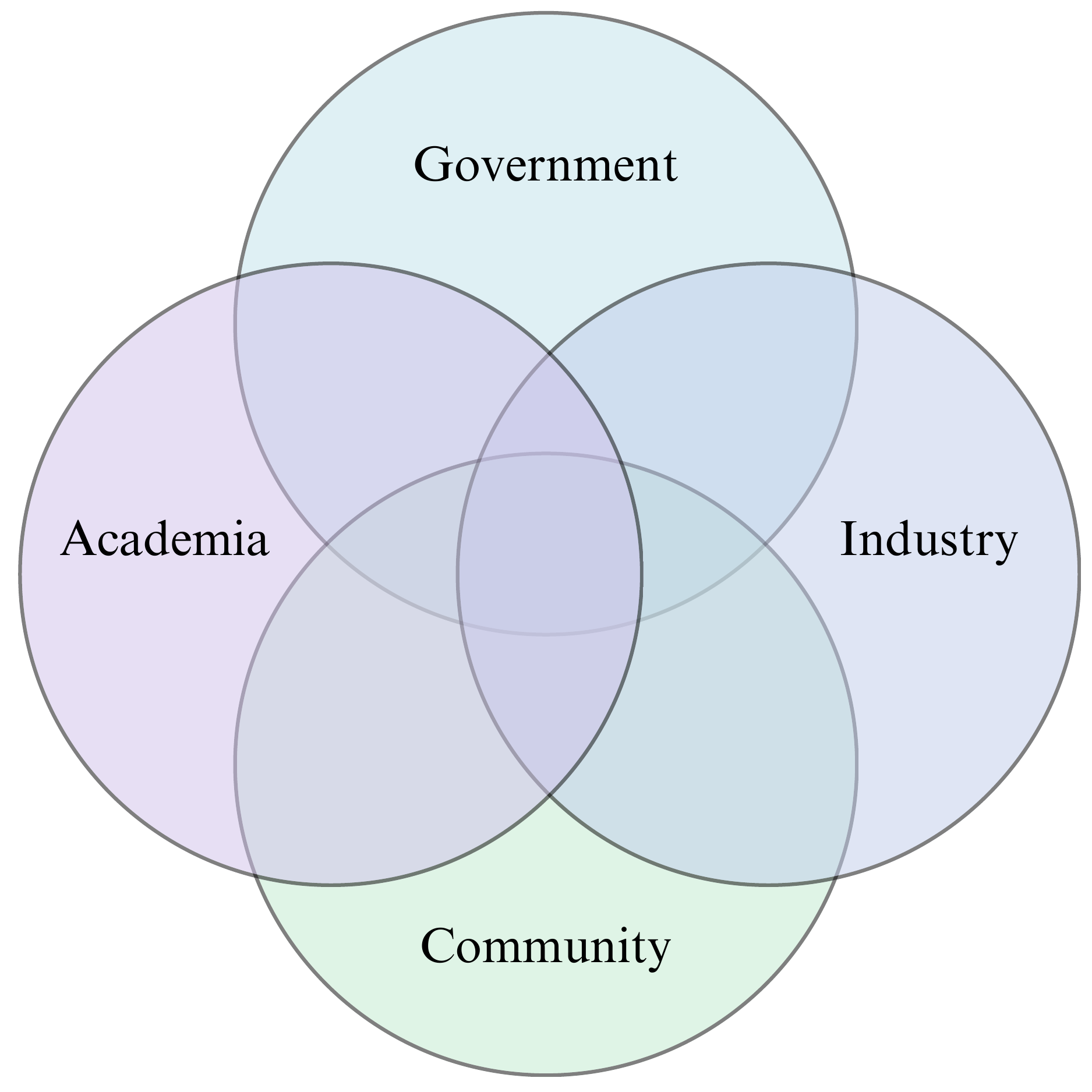}
    \caption{Quadruple helix}
    \label{fig: quadruple}
    \end{figure}

Considering SDDI topics, we further believe that focusing solely on academia and industry for knowledge transfer is insufficient.
SDDI initiatives are widespread beyond academia and industry---and fragmented.
For example, government bodies pass legislation related to diversity and representation\footnote{For instance, the European Accessibility Act and the US Americans with Disabilities Act (ADA), Section 508}, and societal groups promote specific aspects of diversity\footnote{For instance, ACM-W, which is "supporting, celebrating, and advocating for Women in Computing".} and techniques toward a more inclusive society. 
Thus, knowledge transfer of SDDI research is essential for each of the different actors in the quadruple helix (Figure \ref{fig: quadruple}): society, academia, government, and industry~\cite{carayannis2009mode}. We have chosen the quadruple helix approach to visualize all the actors in our field: In the Quadruple Helix model, society refers to citizens, media, cultural organizations, and non-profit organizations. In the software engineering field, these kinds of actors could be media outlets such as TechCrunch and Wired or non-profit communities and organizations such as 'Girls Who Code' or Czechitas~\cite{Buhnova2024}. In business, they refer to software engineering companies, and in government, they refer to the public policy around technology or public funding for software companies and non-profit organizations. In academia, it refers to universities, research centres and conferences. 

As a result of this spread and fragmentation, encouraging actions and initiatives may overlap and remain invisible to other actors. More efficient knowledge transfer between different actors could benefit all, particularly as some technology companies are reducing their DEI teams ~\cite{burnett2024prevent, alfonseca2023corporate}. By engaging actors from academia, industry, government, and society, we could promote more efficient inclusive mentorship practices that foster the inclusion of women and minorities in professional and OSS contexts~\cite{jacobs2024mentorship, hyrynsalmi2019role}. Up-to-date knowledge from different sectors can also prevent situations in which well-meant policies can lead to unintended negative consequences or tokenism~\cite{burnett2024prevent}.
Therefore, we argue that research that considers SDDI in SE needs to engage with all four areas in the quadruple helix. To aim for an even more ambitious knowledge transfer level, we might even consider including the 'environment' in knowledge transfer, i.e., the sustainability aspect. This leads to a 'Quintuple helix', which is especially focused on enabling sustainable and environmentally conscious innovation, as the quadruple helix emphasizes innovation driven by collaboration between the private sector, government, academia, and civil society \cite{carayannis2009mode, carayannis2012quintuple}.

Finally, re-considering the direction of knowledge transfer is important for research related to SDDI topics.
In addition to transferring research results from academia to industry, society, and governments, researchers need to improve their knowledge of initiatives and the results obtained in the broader societal context.
This also relates to the \emph{Methodology and Metrics} theme, as methods that include stakeholders could be beneficial to reach this goal, e.g., participatory research or co-creation. 
Overall, we summarize our discussion on this theme in the following research goal:\greybox{
\textbf{Research Goal:} Engage in knowledge transfer among the quadruple helix of industry, academia, government, and society, considering both transfer to and from academia to the remaining actors.
}

\subsection{Connections to Socioeconomic Understanding}

Given the spread of IT across nearly all aspects of human life, we have to acknowledge that socioeconomics has an impact on our field.
Diversity and inclusion are not merely SE problems. They are projections of much broader socioeconomic problems, which have been studied in research from multiple social science disciplines, e.g., sociology, anthropology, and education. Compared to SE research, they have developed socioeconomic understandings of these issues at much larger scale \cite{babbie2020practice}.
Such large-scale understandings and small-scale context-focused research in SE could complement each other. By connecting with these societal scale understandings, we might better distinguish the unique problems in SE and common social problems, position ourselves and our research in the full social spectrum, understand diversity and inclusion problems' socioeconomic roots, inspire novel interventions, and coordinate to tackle diversity and inclusion problems in SE as a part of global social forces. From a micro-perspective at the individual level, socioeconomic understandings could remind us to seek solutions for the undesirable workplace scenario described in Scenario S3.  

Social scientists have established such societal scale understandings mostly by tracking the socioeconomic dynamics over relatively long periods, represented by the major multi-wave, nationwide, or international surveys, such as the General Social Survey\footnote{\url{https://gss.norc.org/}}, World Value Survey\footnote{\url{https://www.worldvaluessurvey.org/}}, and the Bureau of Labor Statistics\footnote{\url{https://www.bls.gov/cps/lfcharacteristics.htm}}. The results of these surveys could be integrated with our research through a number of different ways. The results could be used in quantitative analysis to identify potential relationships, e.g., the overall labour market dynamics and women's involvement in SE, in which data from the Bureau of Labor Statistics could be used. They might also provide contextual information in qualitative inquiries, e.g., the World Value Survey may help SE researchers interpret qualitative data about the differences in women's participation between the United States and China. Besides, when designing and delivering educational materials, these socioeconomic understandings can offer unique insights to help us better understand the audience.  

\begin{table}
\centering
\caption{Connecting Large Scale Socioeconomic Understandings: Benefits, Harms, and Research Directions}
\label{largescale}
\footnotesize
\begin{tabular}{|p{0.3\linewidth}|p{0.3\linewidth}|p{0.3\linewidth}|}
\hline
\textbf{Benefits} & \textbf{Harms} & \textbf{Research Directions} \\
\hline
\begin{itemize}[leftmargin=5pt]
    \item Better and more comprehensive understandings of diversity and inclusion issues within and beyond the software engineering industry
    \item Coordinated effort to address larger problems
    \item Potential impacts across traditional discipline boundaries
\end{itemize} & 
\begin{itemize}[leftmargin=5pt]
    \item Incorrectly attributing SE-specific problems to general socioeconomic problems
    \item Taking a passive attitude to wait for socioeconomic changes
    \item Ignoring the research results from other disciplines
    \item Using socioeconomic factors as an excuse for the inaction
\end{itemize} & 
\begin{itemize}[leftmargin=5pt]
    \item Exploring forms of connecting socioeconomic understandings with SE research in diversity and inclusion
    \item Developing customized interventions considering different groups' socioeconomic backgrounds 
    \item Coordinating with social scientists to address national/international diversity and inclusion issues
    \item Preparing for the potential socioeconomic changes resulting from recent progresses in generative AI
\end{itemize} \\
\hline
\end{tabular}
\label{tab-Socioeconomic}
\end{table}

Connecting SE to a large-scale socioeconomic understanding requires intensive interdisciplinary collaboration between SE researchers and social scientists. However, we seldom see SE research published in social science venues, and vice versa. In the 2030s, we expect there will be a significant increase in interactions between both sides. Although such connections may be beneficial, they are not without risk. In particular, it may lead to some inertia before certain socioeconomic conditions improve or to some excuses for inaction in our industry. Table \ref{tab-Socioeconomic} summarizes the potential benefits, harms, and research directions. 
We formulate the following research goal for this theme:

\greybox{
\textbf{Research Goal:} Understand and address challenges inhibiting software engineers from disadvantaged socioeconomic backgrounds.
}

\section{Discussion}
In the following, we discuss the four research themes jointly towards an agenda of SDDI in SE.
We then add a brief discussion of how recent changes in AI and education relate to the outlined agenda.

\subsection{An Agenda of SDDI}
Our four research themes and their research goals jointly provide an actionable agenda, bridging gaps in SDDI to make the two desirable ``utopian'' scenarios possible and avoid the two undesirable ``dystopian'' scenarios.
First, appropriate research on \emph{methodologies and metrics} must be developed to effectively analyze various aspects of SDDI in SE.
Specifically, two concrete aspects of SDDI that we believe are of particular importance are the \emph{intersectionality} of software engineers and the connection of \emph{large-scale socioeconomic understandings} to SE practice.
These two aspects need to be studied and understood more thoroughly to suggest appropriate SDDI initiatives and actionable principles for SE practitioners.
Finally, SE research connected to SDDI will not impact education or practice without successful \emph{knowledge transfer}.
Given the relevance of societal and governmental initiatives to SDDI, transferring to and from these groups requires dedicated focus.
In summary, our research goals for the four themes are as follows:
\begin{itemize}
    \item Methodologies and Metrics: Develop methodologies and metrics to effectively analyze diversity and inclusion in software engineering, making use of mixed methods and online community data.
    \item Intersectionality: Understand challenges and motivate solutions to support developers who identify with multiple marginalized groups.
    \item Knowledge Transfer: Engage in knowledge transfer among the quadruple helix of industry, academia, government, and society, considering both transfer to and from academia to the remaining actors.
    \item Connections to Socioeconomic Understanding: Understand and address challenges inhibiting software engineers from disadvantaged socioeconomic backgrounds.
\end{itemize}

Research on SDDI typically deals with marginalized groups.
As such, this type of research requires a constant focus on maximizing benefits while minimizing harms, especially to vulnerable groups.
The contrast between the desired and the undesired state in our scenarios highlights this fine balance in an extreme way. Metrics, diversity dimensions, and socioeconomic understanding can and have been used both to the benefit and to the disadvantage of various societal groups.
As a research community, we must strike this balance in a responsible way.

\subsection{The Impact of AI on SDDI}

    \begin{figure}
    \centering
    \includegraphics[width=\textwidth]{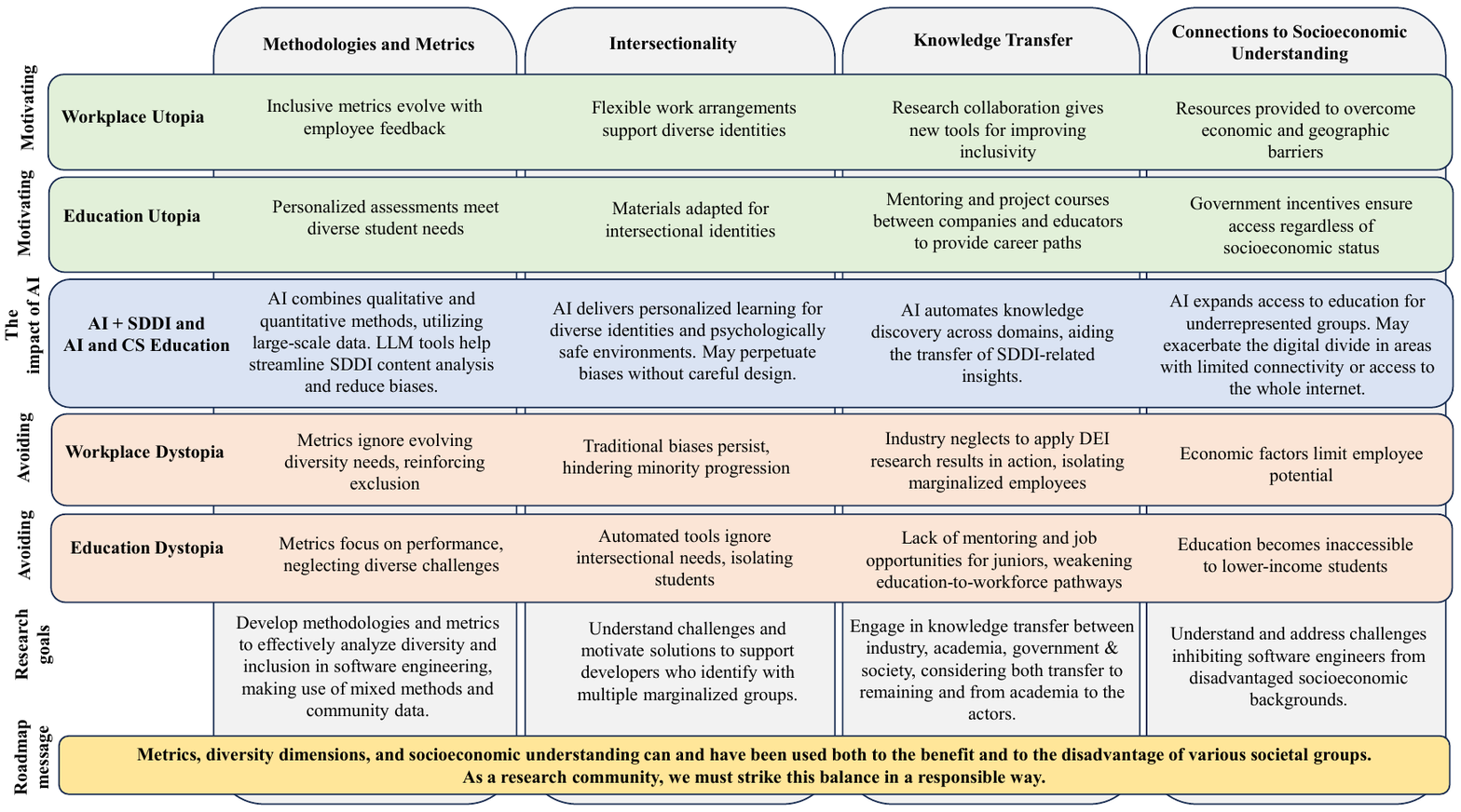}
    \caption{SE 2030 Roadmap for SDDI}
    \label{fig: summary}
    \end{figure}

Although our main themes were methodologies and metrics, intersectionality, large-scale socioeconomic data, and knowledge
transfer, two cross-cutting themes emerged from the group work at the original workshop, namely the two AI-related topics AI \& SDDI and AI \& CS education. Therefore, we add to our roadmap for 2030 for SDDI a particular discussion on AI (Figure \ref{fig: summary}). 

Recent progress in generative AI, exemplified by the advent of LLMs and the burgeoning trend of computing as a general education, holds transformative potential for SDDI research. These advances are poised to revolutionize every facet of our agenda, sparking new avenues of exploration and understanding across research, industry, and educational contexts to build a more inclusive future in software development.
We should be mindful that AI-based solutions not only inherit traditional \textbf{SDDI challenges}, but also come with new ones.

\emph{Languages:} For example, solutions produced by GitHub Copilot for Chinese prompts were found to be subpar compared to their English and Japanese counterparts~\cite{Koyanagi2024}. This might create obstacles for developers preferring to express themselves in Chinese. Moreover, while AI-generated contents or AI-based translation techniques lower the barriers for non-native English speakers' participation, using they may also lead to potential biases towards those non-native speakers regarding their capabilities \cite{wang2024uncovering}.   

\emph{Cognitive Styles:} Similarly, using ChatGPT-like solutions often involves playful experimentation known as ``tinkering'', a learning strategy more common among men than among women~\cite{beckwith2006tinkering,burnett2016gendermag}.
Interestingly enough, a recent study has found men believing that ChatGPT did not personalize their experience enough and did not allow global customization of its behaviour, which the authors attribute to the insufficient amount of tinkering supported by ChatGPT~\cite{Choudhuri2024How}. 
Tinkering is only one of the aspects of ChatGPT-like solutions where perceptions of individuals of different genders might diverge. For example, women tend to consider the decision-making process of ChatGPT to be less transparent than men~\cite{Choudhuri2024How}.
In contrast, ChatGPT-like solutions could help several groups of developers who might feel more comfortable posing their questions to a machine rather than asking people, for example, neurodivergent developers who often face difficulties in communication~\cite{morris2015understanding}.

\emph{Cultural Aspects:} Moreover, \citeauthor{atari_xue_park_blasi_henrich_2023} have shown that LLM responses to psychological measures are not aligned with large-scale cross-cultural data~\cite{atari_xue_park_blasi_henrich_2023}. They have also shown that the performance of LLMs in cognitive psychological tasks is most similar to that of people from WEIRD societies. 

\emph{Learning:} In a study with more than 3,000 software engineers, \citeauthor{hicks_lee_foster-marks_2024} have found that the ``AI Skill Threat'', that is the struggle with adapting to AI-assisted work, is more common for racially minoritized developers~\cite{hicks_lee_foster-marks_2024}. That group also rated the quality of the output of AI-assisted coding tools significantly lower than other groups~\cite{hicks_lee_foster-marks_2024}.

In summary, although aspects such as gender bias in software design~\cite{DBLP:journals/sigcse/Huff02} and gender difference in human-computer interaction~\cite{DBLP:conf/chi/CzerwinskiTR02, DBLP:conf/chi/TanCR03} and programming environments~\cite{DBLP:conf/vl/BeckwithB04} have been discussed already in the early 2000s, it took years, if not decades, for the broader topic of diversity, equity, and inclusion to reach the main discourse in the software engineering research community.
With this paper, we want to raise awareness that, as motivated above, GenAI inherits traditional SDDI challenges and comes with new ones.
We hope that both researchers and practitioners consider the existing body of knowledge on SDDI topics when developing and improving GenAI models and tools~\cite{sddibook2024}.

Regarding \textbf{methodologies and metrics}, AI may offer the unprecedented capability to bridge existing qualitative and quantitative methods, enabling the development of deep insights about SDDI from a large volume of data, such as online community data. For example, LLMs' automated sentiment and opinion mining features could significantly accelerate the process of analyzing qualitative data to identify SDDI-related content~\cite{lin2024inditag}, and improve the effectiveness in quantifying hard-to-detect implicit biases~\cite{10.1109/ICSE-SEIS.2019.00009, 10.1145/3368089.3409762}. Their multilingual features may ease the process of research focusing on non-WEIRD populations. The survey design and execution process may also be partially automated with AI techniques; for example, LLMs could help summarize related literature, particularly literature outside the SE domain, to identify potential metrics for constructs related to diversity and inclusion. These methodologies and metrics can then be used to inform SE education and industry practices. Meanwhile, we must acknowledge that most AI technologies inherit biases and discrimination from diverse sources \cite{ferrer2021bias}. Thus, SDDI researchers must be cautious and vigilant when integrating AI into their methodological arsenal. 

AI can also play a role in \textbf{intersectionality}, affecting SDDI in computing education and practice. In addition to the methodological benefits mentioned above, AI could play a positive role in SE education involving people with diverse identities. One of the major promises of AI is to provide personalized materials to people of different characteristics \cite{10.1145/3580305.3599572}. Thus, AI could bring individualized learning experiences to individuals of certain intersectionality, such as gender-sensitive, accessible software development learning materials for students (see S2: Desired Scenario in Education). Moreover, conversational agents powered by AI techniques have the potential to create psychologically safe development environments in which individuals of certain intersectionality would not feel embarrassed when interacting with AI. However, researchers must keep in mind that individuals from different identities might interact with AI technologies differently or that they might benefit differently from AI. For instance, studies report developers who identify as female and LGBTQ+ have significantly lower intent to learn and adopt AI-assisted coding platforms, while software engineers of racial minorities have a higher intent to upskill but also more negative perceptions of AI compared to their counterparts~\cite{pluralsight}.

As mentioned above, the \textbf{knowledge transfer} of SDDI topics is essential for different actors in the quadruple helix (society, academia, government, and industry). The key challenge is the invisibility among actors. AI technologies, due to their capability to automate knowledge discovery, may partially mitigate this challenge. LLMs could help automate the cumbersome process of identifying and distilling the widespread knowledge in the interdisciplinary literature and then compile it from fragmented pieces into organized knowledge bases or repositories~\cite{NEURIPS2023_83fc8fab}. Conversational agents powered by AI might facilitate the dissemination of information and knowledge among actors from different quadruple helix to increase the visibility of those actors' efforts. SDDI researchers may create domain-specific LLMs as the interfaces for engaging with key actors from other domains. Meanwhile, the relevant SDDI knowledge could be better integrated into their initiatives, including their effort to make computing education more accessible.    

Generative AI technologies would inevitably change the \textbf{landscape of today's socioeconomic situations}. These technologies might create new disparities that have never been seen before. For example, well-paid professional labor markets such as software development might experience a reduction~\cite{acemoglu2022artificial}. How can we make various minority group members suffer less if this happens? How can we reskill minority group members to participate in future work? How can we avoid further polarization of the labor market? To address these issues, SDDI researchers should collaborate with researchers from other disciplines to closely monitor the socioeconomic dynamics and develop forward-looking solutions. When it comes to educational contexts, AI could help to scale computing education to larger cohorts as the adoption of computing education rises across all disciplines~\cite{CC2020}. In particular, AI can support learning for individuals who may not otherwise be able to receive such education due to their socioeconomic backgrounds---for example, learners from the Global South. However, access to advanced AI leads to a new digital divide~\cite{khowaja2024chatgpt}. While LLMs are almost ubiquitous in high-income economies, reliable Internet access remains a big issue in many low-income economies~\cite{kula2022png}, contributing to the Undesirable Workplace Scenario (S3) described in Section~\ref{sec:scenarios}. In this process, people from minority groups may lose their human anchors, which gives them a sense of belonging that keeps them in the area (see Undesirable Scenario in Education, S4). Hence, SDDI research should not underestimate the potential socioeconomic changes caused by fast-evolving generative AI technologies.

\section{Conclusions}

Software affects almost all areas of modern life, affecting user behaviour, well-being, and decision-making. Software engineers design and develop software applications---yet the diversity of software development teams does not represent the diversity of the global population. To this end, this work presents insights from SE researchers and practitioners on challenges and research opportunities to promote software developer diversity and inclusion (SDDI). We also offer a SE 2030 Roadmap for SDDI to encourage new research approaches for diversity and inclusion in software development. We provide motivating desired (``utopian'') and undesired (``dystopian'') scenarios describing the effects of diversity on SE practice and education in 2030 and discuss ways to promote SDDI through research methodologies, intersectionality, knowledge transfer, and socioeconomic understanding to navigate the changing landscape of software development. We further briefly discuss the potential impact of the recent progress in generative AI and the burgeoning trend of computing as a general education.

The initial themes presented in this paper were developed during a research meeting in 2023.
We picked up these six themes in this paper, four of them discussed as dedicated research themes and the remaining two as a part of the undesirable/desirable scenarios and the discussion.
These themes represent our collective opinions on the key challenges and research opportunities for promoting SDDI.
The prioritization approach used through the 1-2-4-ALL structure, combined with the extensive collective experience of the FSE SE 2030 workshop participants, builds confidence that these themes are indeed the most relevant ones for future research endeavours.
Furthermore, our discussion adds relevant existing research to support the importance of the themes.
However, ultimately an argument could be made for a different focus and/or for different research themes.

\section*{Acknowledgements}

Some initial ideas presented in this paper were originally developed at the 2023 Software Developer Diversity and Inclusion Workshop. We thank the NII Shonan meeting for hosting the workshop and all workshop participants who contributed to the discussions and our original report. Rafael Prikladnicki thanks CNPq in Brazil. 


\bibliographystyle{ACM-Reference-Format}
\balance
\bibliography{software}

\end{document}